\documentstyle[12pt]{article}

\begin{document}
\def\S{ + \!\!\!\!\!\! \supset }

\title{ON  CONTRACTIONS OF CLASSICAL BASIC SUPERALGEBRAS}

\author{N. A. GROMOV, I. V. KOSTYAKOV, V. V. KURATOV \\
 Department of Mathematics, \\ 
Syktyvkar Branch of IMM UrD RAS,  \\
Chernova st., 3a, Syktyvkar, 167982, Russia \\
E-mail: gromov@dm.komisc.ru
}

\maketitle
\begin{abstract}{ We define a class of orthosymplectic 
$osp(m;j|2n;\omega)$ and unitary $sl(m;j|n;\epsilon)$  superalgebras
 which may be obtained from
$osp(m|2n)$ and $sl(m|n)$ by contractions and analytic continuations 
in a similar way as the special linear, orthogonal and the symplectic 
Cayley-Klein algebras are obtained from the corresponding classical ones. 
Casimir operators of Cayley-Klein superalgebras are obtained from 
the corresponding operators of the basic superalgebras.
Contractions of $sl(2|1)$ and $osp(3|2)$ are regarded as an examples. }
\end{abstract}

\section{Introduction}
Since their discovery \cite{1},
\cite{2}, \cite{3} in 1971 the supersymmetry is used in different 
physical theories such as Kaluza--Klein supergravity \cite{W-86},
supersymmetric field theories of the Wess--Zumino type \cite{K-75},
massless higher-spin field theories \cite{Vas-90}. Recently the secret
theory \cite{B-96} (or  S-theory) that includes superstring theory and its
super p-brane and D-brane \cite{BIK}  generalizations was discussed.
All these theories are build algebraically with the help of some
superalgebra in their base. In this work we wish to present a wide
class of Cayley--Klein (CK) superalgebras which may be used for
constructions of different sypersymmetric models.

For an ordinary Lie groups (or algebras) the title CK was initially used
for the short name of the set of a motion groups of a spaces of constant
curvature. It is well known that there are $3^n$  n-dimensional 
spaces of constant curvature and their motion groups may be obtained
from the orthogonal group $SO(n+1)$ with the help of contractions 
and analytical continuations \cite{NG}. Later the notion CK was extend
to the case of unitary and symplectic groups (algebras) \cite{JMP}.
The typical (and attractive) property of CK groups is that all of them
are depend on the same number of independent parameters as the
corresponding simple classical group. On the level of Lie algebras 
this means that all CK algebras of the same type have the equal 
dimensions. A basic superalgebras include a simple algebras 
as an even subalgebras, so it looks quite natural to introduce 
a new class of superalgebras with CK algebras as an even subalgebras.

A superalgebra as an algebraic structure contain (as compared with
Lie algebra) a new additional operation, namely, $Z_2$-grading.
So under contraction of superalgebra this $Z_2$-grading must be
conserved. To our knowledge contraction of orthosymplectic
superalgebra  to the superkinematics was first regarded in \cite{Rem}.
The detailed investigation a class of contraction of $osp(1|2)$
and $osp(1|4)$ to the kinematical Poincar$\acute{e}$ and Galilei
superalgebras was made in \cite{Val-99}. Contraction  of unitary
superalgebra $Gsu(2)=sl(2|1)$ as well as their representations
was described  in \cite{Pat}. Later the notion of contraction 
was generalized  \cite{MP} to the case of Lie algebra 
with an arbitrary finite grading group and is known 
as graded contractions. Nevertheless the particular case of the simplest
$Z_2$-grading deserve an independent interest. 
The preliminary results was reported in \cite{GKK}.

The paper is organized as follows. In section 2 the orthogonal,
symplectic and special linear CK groups and algebras are briefly remind.
Section 3 is devoted to the orthosymplectic CK superalgebras.
CK unitary superalgebras are regarded in section 4. Casimir operators
of the CK unitary and orthosymplectic superalgebras are described
in section 5.

\section{ Orthogonal, symplectic and special linear Cayley-Klein algebras}

 Special linear $sl(m),$
orthogonal $so(m)$ and symplectic $sp(2n)$ algebras are even
subalgebras of classical basic superalgebras. 
On the other hand all of them 
may be contracted and analytically continued to the set of 
CK  algebras. Lie groups and algebras are in close relations.
CK group $SO(m;j)$ is defined as the set of transformations
of vector space ${\bf R}_m(j),$ which preserve the quadratic form
$x^2(j)=x^t(j)x(j)
=x_1^2+\sum_{k=2}^{m}(1,k)^2x^2_k, $
where $ (i,k)=\prod^{\max(i,k)-1}_{p=\min(i,k)}j_p,
\, (i,i)=1, $
each parameter $j_k=1,\iota_k,i,$
where $\iota_k$ are nilpotent $ \iota^2_k=0,$ commutative
$\iota_k\iota_p=\iota_p\iota_k \neq 0$
generators of Pimenov algebra ${\bf}P(\iota).$
Cartesian components of vector $x(j)\in {\bf R}_m(j)$
are $x^t(j)=(x_1,j_1x_2, \ldots ,(1,m)x_m)^t, $
as it is easily follows from $x^2(j).$
For $m\times m$ matrix $g(j) \in SO(m;j)$ the transformation
$g(j): {\bf R}_m(j) \rightarrow {\bf R}_m(j)$ means that the vector
$x'(j)=g(j)x(j)$ has exactly the same distribution of
parameters $j$ among its components as $x(j).$
This requirement give an opportunity to obtain the
distribution of parameters $j$ among elements of matrix $g(j),$
i.e. to build the fundamental representation of CK group
$SO(m;j)$ starting from the quadratic form. It is remarkable that the
same distribution of the parameters $j$ is hold for CK Lie
algebra $so(m;j),$ namely
$A_{ik}=(i,k)a_{ik},$ for $A \in so(m;j).$

The set of transformations
$ L(j): {\bf R}_m(j)\to {\bf R}_m(j) $
with the property 
$ \det L(j)=1 $ form CK special linear group $ SL(m;j)$ and the
corresponding CK algebras $ sl(m;j)$ are given by the
$m \times m $ matricies $l(j),$  tr $ l(j)=0.$
Let us stress that in Cartesian basis all matricies from
$ SL(m;j), SO(m;j), sl(m;j), so(m;j) $ have
identical distribution of parameters $j$ among its elements, i.e.
they are of the same type as the matricies with elements from
Pimenov algebra $P(j).$

CK symplectic group $Sp(2n;\omega)$ is defined as the set of
transformations of ${\bf R}_n(\omega) \times {\bf R}_n(\omega),$
which preserve the bilinear form
$S(\omega)=S_1+
\sum_{k=2}^{n}[1,k]^2S_k,$ where $S_k(y,z)=y_kz_{n+k}-y_{n+k}z_k, \,
[i,k]=\prod^{\max(i,k)-1}_{p=\min(i,k)}
\omega_k, \, [i,i]=1, \,
\omega_k=1,\xi_k,i, \, \xi^2_k=0, \,
\xi_k\xi_p=\xi_p\xi_k.$
The distribution of parameters $\omega_k$ among matrix elements
of the fundamental representation
$M(\omega)=\left( \begin{array}{cc}
 H(\omega) & E(\omega) \cr
F(\omega) & -H^t(\omega) \end{array} \right)$
of the CK symplectic algebra $sp(2n;\omega)$ may be obtained as for
orthogonal CK algebras and is as follows:
$B_{ik}=[i,k]b_{ik},$ where $B=H(\omega),E(\omega),F(\omega).$

\section{ Orthosymplectic superalgebras $osp(m;j|2n;\omega)$}

 Let $e_{IJ} \in M_{m+2n}$ satisfying $(e_{IJ})_{KL}=\delta_{IK}\delta_{JL}$
are elementary matrices. One defines the following
graded matrix \cite{Fra}
\begin{equation}
G=\left (
\begin{array}{c|c}
 I_m   &   0   \cr \hline
0  &   0 \quad  I_n   \cr
   &  -I_n \quad 0
\end{array}
\right )
\label{0}
\end{equation}
where $I_m,I_n$ are identity matrices. Let $i,j,\ldots=1,\ldots ,m, \,
\bar i,\bar j=m+1, \ldots, m+2n.$
The generators of the orthosymplectic superalgebra $osp(m|2n)$ are given by
$$
E_{ij}=-E_{ji}=\sum_{k}(G_{ik}e_{kj}-G_{jk}e_{ki}),\;\;
E_{\bar i\bar j}=E_{\bar j\bar i}=\sum_{\bar k}
(G_{\bar i\bar k}e_{\bar k\bar j}+
G_{\bar j\bar k}e_{\bar k\bar i}),\;\;
$$
\begin{equation}
E_{i\bar j}=E_{\bar ji}=\sum_{k}G_{ik}e_{k\bar j}+
\sum_{\bar k}G_{\bar j\bar k}e_{\bar ki},
\label{1}
\end{equation}
where the even (bosonic) $E_{ij}$ generate the  $so(m)$ part,
the even (bosonic)
 $E_{\bar i\bar j}$ generate the $sp(2n)$ part and the
 rest $E_{i\bar j}$ are the  odd
 (fermionic) generators of superalgebra. They satisfy
 the following (super) commutation relations
$$
[E_{ij},E_{kl}]=G_{jk}E_{il}+G_{il}E_{jk}-G_{ik}E_{jl}-G_{jl}E_{ik},\;\;
$$
$$
[E_{\bar i\bar j},E_{\bar k\bar l}]=-G_{\bar j\bar k}E_{\bar i\bar l}-
G_{\bar i\bar l}E_{\bar j\bar k}-G_{\bar j\bar l}E_{\bar i\bar k}-
G_{\bar i\bar k}E_{\bar j\bar l},
$$
$$
[E_{ij},E_{k\bar l}]=G_{jk}E_{i\bar l}-
G_{ik}E_{j\bar l},\;\;
[E_{i\bar j},E_{\bar k\bar l}]=-G_{\bar j\bar k}E_{i\bar l}-
G_{\bar j\bar l}E_{i\bar k},
$$
\begin{equation}
[E_{ij},E_{\bar k\bar l}]=0, \quad
\{E_{i\bar j},E_{k\bar l}\}=
G_{ik}E_{\bar j\bar l}-
G_{\bar j\bar l}E_{ik}.
\label{2}
\end{equation}

In the matrix form
$
osp(m|2n)=\{M \in M_{m+2n}|M^{st}G+GM=0\}.
$
If the matrix $M$ has the following form:
$
M=\sum_{i,j}a_{ij}E_{ij} + \sum_{\bar i,\bar j}b_{\bar i\bar j}
E_{\bar i\bar j} + \sum_{i\bar j}\mu_{i\bar j}E_{i\bar j},
$
with $a_{ij}, b_{\bar i \bar j}\in $ {\bf R} or {\bf C} and
$\mu_{i\bar j}$ as the odd nilpotent elements of
Grassmann algebra: $\mu^2_{i\bar j}=0,\,
\mu_{i\bar j}\mu_{i'\bar j'}=-\mu_{i'\bar j'}\mu_{i\bar j},$
then the corresponding supergroup $Osp(m|2n)$ is obtained by the
exponential map $ {\cal M}=\exp M $ and act on (super)vector space
by matrix multiplication ${\cal X}'={\cal M}{\cal X},$ where
${\cal X}^t=(x|\theta)^t,$ $x$
is a $n$--dimentsional even vector and $\theta$ is
a $2m$--dimensional odd vector with odd Grassmann elements.
The form $inv=\sum^m_{i=1}x^2_i+2\sum^n_{k=1}
\theta_{+k}\theta_{-k}=x^2+2\theta^2$
is invariant under this action of orthosymplectic supergroup.

We shall define CK orthosymplectic superalgebras starting with
the invariant form
\begin{equation}
inv(j;\omega) = u^2\sum^m_{k=1}(1,k)^2x^2_k + 
2v^2\sum^{m+n}_{\bar k=m+1}[1,\hat {\bar k}]^2
\theta_{\hat {\bar k}}\theta_{-\hat {\bar k}}\equiv 
u^2x^2(j) + 2v^2\theta^2(\omega),
\label{4}
\end{equation}
$\hat {\bar k}=\bar k-m,$ when $ \bar k=m+1,\ldots ,m+n $ and
$\hat {\bar k}=\bar k-m-n,$ when  $ \bar k=m+n+1, \ldots ,m+2n,$
which is the natural unification of CK orthogonal and
 symplectic forms. The distributions of contraction parameters
$j,\omega $ among matrix elements of the fundamental
representation of $osp(m;j|2n;\omega)$
and the transformations of the generators (\ref{1})
 are obtained in a standart CK manner and are as follows:
\begin{equation}
E_{ik}=(i,k)E^*_{ik}, \;\;
E_{\bar i\bar k}=[\hat {\bar i},\hat {\bar k}]E^*_{\bar i\bar k},\;\;
E_{i\bar k}=u(1,i)v[1,\hat {\bar k}]E^*_{i\bar k},
\label{5}
\end{equation}
where $E^*$ are generators (\ref{1}) of the starting superalgebra
$osp(m|2n).$ The transformed generators are subject of the
(super) commutation relations:
$$
[E_{ij},E_{kl}] = (i,j)(k,l) \left (
{{G_{jk}E_{il}} \over {(i,l)}} + {{G_{il}E_{jk}} \over {(j,k)}} -
{{G_{ik}E_{jl}} \over {(j,l)}} - {{G_{jl}E_{ik}} \over {(i,k)}} \right ),
$$

 $$
[E_{\bar{i}\bar{j}},E_{\bar{k}\bar{l}}] =
-[\hat{\bar{i}},\hat{\bar{j}}][\hat{\bar{k}},\hat{\bar{l}}] \left (
{{G_{\bar{j}\bar{k}}E_{\bar{i}\bar{l}}} \over 
{[\hat{\bar{i}},\hat{\bar{l}}]}} +
{{G_{\bar{i}\bar{l}}E_{\bar{j}\bar{k}}} \over 
{[\hat{\bar{j}},\hat{\bar{k}}]}} + 
{{G_{\bar{i}\bar{k}}E_{\bar{j}\bar{l}}} \over 
{[\hat{\bar{j}},\hat{\bar{l}}]}} +
{{G_{\bar{j}\bar{l}}E_{\bar{i}\bar{k}}} \over 
{[\hat{\bar{i}},\hat{\bar{k}}]}} \right ),
$$
$$
[E_{ij},E_{\bar{k}\bar{l}}] = 0, \quad
[E_{ij},E_{k\bar{l}}] =
(i,j)(1,k) \left (
{{G_{jk}E_{i\bar{l}}} \over {(1,i)}} -
{{G_{ik}E_{j\bar{l}}} \over {(1,j)}} \right ),
$$
 $$
[E_{i\bar{j}},E_{\bar{k}\bar{l}}] =
-[1,\hat{\bar{j}}][\hat{\bar{k}},\hat{\bar{l}}] \left (
{{G_{\bar{j}\bar{k}}E_{i\bar{l}}} \over {[1,\hat{\bar{l}}]}} +
{{G_{\bar{j}\bar{l}}E_{i\bar{k}}} \over {[1,\hat{\bar{k}}]}} \right ),
$$
\begin{equation}
\{E_{i\bar{j}},E_{k\bar{l}}\} =u^2v^2
(1,i)[1,\hat{\bar{j}}](1,k)[1,\hat{\bar{l}}] \left (
{{G_{ik}E_{\bar{j}\bar{l}}} \over {[\hat{\bar{j}},\hat{\bar{l}}]}} -
{{G_{\bar{j}\bar{l}}E_{ik}} \over {(i,k)}} \right ).
\label{6}
\end{equation}
For $u=\iota$ or $v=\iota, \iota^2=0$ superalgebra $osp(m|2n)$
is contracted to inhomogeneous superalgebra, which is semidirect sum
$ \{E_{i\bar{j}}\} \S (so(m) \bigoplus sp(2n)),$
with all anticommutators of the odd  generators equal to zero
$\{E_{i\bar{j}},E_{k\bar{p}} \} = 0.$

\subsection{Example: CK contractions of $osp(3|2)$}

This superalgebra has $so(3)$ as even subalgebra therefore
their contractions to the kinematical $(1+1)$
Poincare, Newton and Galilei superalgebras may be
fulfilled according to general CK scheme of the first section.
But unlike of two odd generators of $osp(1|2)$ the superalgebra
$osp(3|2)$ has six odd generators. In the basis 
$X_{ik}=E_{ki}, \, k,i=1,2,3, \, F=\displaystyle{\frac{1}{2}}E_{44}, \,
E=-\displaystyle{\frac{1}{2}}E_{55}, \,
H=-E_{45}, \, Q_k=E_{k4}, \, Q_{-k}=E_{k5}$
the generators are affected by the contraction
coefficients $j_1,j_2$ in the following way
\begin{equation}
X_{ik}\to (i,k)X_{ik}, \quad Q_{\pm k}\to (1,k)Q_{\pm k}
\label{12}
\end{equation}
and $H,F,E $ are remained unchanged.
Then superalgebra $osp(3;j|2)$ is given by
$$
[X_{12},X_{13}]=j_1^2X_{23}, \quad [X_{13},X_{23}]=j_2^2X_{12}, \quad
[X_{23},X_{12}]=X_{13},
$$
$$
[H,E]=2E, \quad [H,F]=-2F, \quad [E,F]=H,
$$
$$
[X_{ik},Q_{\pm i}]=Q_{\pm k}, \quad
[X_{ik},Q_{\pm k}]=-(i,k)^2Q^2_{\pm i}, \;\;  i<k,
$$
$$
[H,Q_{\pm k}]=\mp Q_{\pm k}, \quad
[E,Q_k]=-Q_{-k}, \quad [F,Q_{-k}]=-Q_k,
$$
$$
\{Q_k,Q_k\}=(1,k)^2F, \quad \{Q_{-k},Q_{-k}\}=-(1,k)^2E,
$$
\begin{equation}
\{Q_k,Q_{-k}\}=-(1,k)^2H, \quad \{Q_{\pm i},Q_{\mp k}\}=\pm (1,k)^2X_{ik}.
\label{12-}
\end{equation}
The non-minimal Poincare superalgebra is obtained for
$j_1=\iota_1, \, j_2=i $ and has the structure
of the semidirect sum
$T \S (\{X_{23}\}\oplus osp(1|2)),$ where abelian
$T=\{X_{12},X_{13},Q_{\pm 2},Q_{\pm 3}\}$ and
$osp(1|2)=\{H,E,F,Q_{\pm 1}\}.$ The Newton superalgebra
$osp(3;\iota_2|2)=T_2 \S osp(2|2),$ where
$T_2=\{X_{13},X_{23},Q_{\pm 3}\} $ and $osp(2|2) $ is generated by
$X_{12},H,E,F,Q_{\pm 1},Q_{\pm 2}.$
Finally the non-minimal Galilei superalgebra may be presented as
semidirect sums
$osp(3;\iota_1,\iota_2|2)=(T \S\{X_{23}\})\S osp(1|2)=
T \S (\{X_{23}\}\oplus osp(1|2)).$

\section{Unitary superalgebras $sl(m;j|n;\epsilon)$}

The superalgebras $sl(m|n)$ can be generated as a matrix superalgebras
by taking matrices of the form \cite{Fra}
\begin{equation}
M= \left(
\begin{array}{cc}
X_{mm} & T_{mn} \cr
T_{nm} & X_{nn}  \cr
\end{array}
\right )
\end{equation}
where $X_{mm}$ and $X_{nn}$ are $gl(m)$ and $gl(n)$ matrices,
$T_{mn}$ and $T_{nm}$ are $m \times n$ and $n \times m$ matrices
respectively, with the supertrace condition
\begin{equation}
{\rm str}(M)={\rm tr}(X_{mm}) -{\rm tr}(X_{nn})=0.
\label{str}
\end{equation}
This matrix superalgebra is the set of transformations of the 
superspace  with $m$ even coordinates 
$x_1,\ldots,x_m$ and $n$ odd ones $\theta_1,\ldots,\theta_n$.

A basis of  superalgebra $sl(m|n)$ can be constructed as follows.
Define the $(m+n)^2-1$ generators
\begin{eqnarray}
E_{ij}=e_{ij}-{1 \over m-n} \delta_{ij}\left (\sum_{k=1}^{m}e_{kk}+
\sum_{\bar{k}=m+1}^{m+n}e_{\bar{k}\bar{k}}\right),
& \qquad &E_{i\bar{j}}= e_{i\bar{j}}, \cr
E_{\bar{i}\bar{j}}=e_{\bar{i}\bar{j}}+{1 \over m-n} 
\delta_{\bar{i}\bar{j}}\left(\sum_{k=1}^{m} e_{kk}+
\sum_{\bar{k}=m+1}^{m+n} e_{\bar{k}\bar{k}}\right),
& \qquad & E_{\bar{i}j}= e_{\bar{i}j},
\end{eqnarray}
where the indices $i,j,\ldots$ run from $1$ to $m$
and $\bar{i},\bar{j},\ldots$  from $m+1$ to $m+n$.
The generators of $sl(m|n)$ in the Cartan-Weyl basis are given by
\begin{eqnarray}
H_i &=& E_{ij}-E_{i+1, j+1}, \quad  \ 1 \leq i \leq m-1, \nonumber \\ 
H_{\bar{i}} &=& E_{\bar{i}\bar{i}}-E_{\bar{i}+1, \bar{i}+1}, \quad \
\ m+1 \leq \bar{i} \leq m+n-1, \nonumber \\
H_m &=& E_{mm}+E_{m+1, m+1}, \nonumber \\
E_{ij} & & {\rm for} \ \ sl(m),  \qquad 
E_{\bar{i}\bar{j}} \qquad {\rm for} \quad sl(n), \nonumber  \\
E_{i\bar{j}} & & {\rm and} \qquad E_{\bar{i}j} \qquad  
{\rm for \ the \ odd \ part }
\label{gn}
\end{eqnarray}
and their commutation relations  looked as
\begin{eqnarray}
 [H_I,H_J] & = & 0, \cr
 [H_K,E_{IJ}] & = & \delta_{IK}E_{KJ}-\delta_{I,K+1}E_{K+1,J}-
\delta_{KJ}E_{IK}+\delta_{K+1,J}E_{I,K+1},   \; (K\neq m),  \cr
 [H_m,E_{IJ}]&=&\delta_{Im}E_{mJ}-\delta_{I,m+1}E_{m+1,J}-
\delta_{mJ}E_{Im}+\delta_{m+1,J}E_{I,m+1},    \cr
[E_{IJ},E_{KL}] & = & \delta_{JK}E_{IL}-\delta_{IL}E_{KJ} \ \qquad
{\rm for}\  E_{IJ} \ {\rm and} \ E_{KL} \ {\rm even},  \cr
[E_{IJ},E_{KL}] & = &\delta_{JK}E_{IL}-\delta_{IL}E_{KJ} \ \qquad
{\rm for}\  E_{IJ} \ {\rm even} \ {\rm and} \ E_{KL} \ {\rm odd},   \cr
\{E_{IJ},E_{KL}\} & = & \delta_{JK}E_{IL}+\delta_{IL}E_{KJ} \ \qquad
{\rm for}\  E_{IJ} \ {\rm and} \ E_{KL} \ {\rm odd}. 
\label{st}
\end{eqnarray}

CK special linear (or unitary) superalgebras
$sl(m;j|n;\epsilon) $ are consistent with the transformations of
(super) vectors
\begin{equation}
{\cal X}^t(j,\epsilon)=(x_1,j_1x_2,\ldots ,(1,m)x_m|\nu(x_{m+1},
\epsilon_1x_{m+2},\ldots ,[1,n]x_{m+n}))^t,
\label{sv}
\end{equation}
where the odd components are denote as
$
x_{m+1}=\theta_1,\ldots, x_{m+n}=\theta_{n}
$ 
and
$ 
\hat{\bar i}={\bar i}-m, \,
\hat{\bar k}={\bar k}-m=1,\ldots ,n,\,
[\hat{\bar i},\hat{\bar k}]=
\prod^{\max(\hat{\bar i},\hat{\bar k})-1}_{l=\min(\hat{\bar i},\hat{\bar k})}
\epsilon_l, \, \epsilon_l=1,\xi_l,i, \,
\xi^2_l=0, \,
\xi_l\xi_p=\xi_p\xi_l \neq 0. $
The components of ${\cal X}(j;\epsilon) $ are choosen in such a way
that the contraction parameters $\epsilon_l $ of the odd
components were independent of the contraction parameters
$j_l $  of the even ones. The transformations 
of the standart generators (\ref{gn}) (marked with star) 
 of the special linear superalgebra $ sl(m|n) $ to
the generators  of $sl(m;j|n,\epsilon)$ are given by
$$
H_I=H^*_I, \; E_{ij}=(i,j)E^*_{ij}, \;
E_{\bar i\bar j}=[\hat{\bar i},\hat{\bar j}]E^*_{\bar i\bar j}, \;
i \neq j, \; \bar i \neq \bar j,
$$
\begin{equation}
E_{i\bar j}=\nu(1,i)[1,\hat{\bar j}]E^*_{i\bar j}, \quad
E_{\bar ij}=\nu(1,j)[1,\hat{\bar i}]E^*_{\bar ij}.
\label{5.1+}
\end{equation}

Nonzero commutators and anticommutators are easily obtained from the
corresponding commutation relations (\ref{st}) of the initial
superalgebra $ sl(m|n) $ in the form
$$
[H_K,E_{IJ}] = \delta_{IK}E_{KJ}-\delta_{I,K+1}E_{K+1,J}-
\delta_{KJ}E_{IK}+\delta_{K+1,J}E_{I,K+1},
$$
$$
[E_{ij},E_{jl}]=\left\{
\begin{array}{ll}
E_{il}, &i<j<l,\; l<j<i,\; l\neq i, \\
(l,j)^2E_{il}, & i<l<j \ {\rm or} \ j<l<i, \\
(i,j)^2E_{il}, & l<i<j \ {\rm or} \ j<i<i, 
\end{array} \right.
$$
$$
[E_{ij},E_{kj}]=\left\{
\begin{array}{ll}
-E_{kj}, &k<i<j,\; j<i<k,\; k\neq j, \\
-(i,j)^2E_{kj}, & i<j<k {\ \rm or \ }  k<j<i, \\
-(i,k)^2E_{kj}, & i<k<j {\ \rm or \ } j<k<i, 
\end{array} 
\right.
$$
$$
[E_{ij},E_{ji}]=(i,j)^2(E_{ii}-E_{jj}),
$$
$$
[E_{\bar{i}\bar{j}},E_{\bar{j}\bar{l}}]=\left\{
\begin{array}{ll}
E_{\bar{i}\bar{l}}, & 
\bar{i} < \bar{j} < \bar{l},\; \bar{l} < \bar{j} < \bar{i},\; 
\bar{l}\neq \bar{i}, \cr

[\hat{\bar{l}},\hat{\bar{j}}]^2  E_{\bar{i}\bar{l}}, & 
\bar{i}< \bar{l}< \bar{j} {\ \rm or \ } \bar{j} < \bar{l} < \bar{i}, \cr
[\hat{\bar{i}},\hat{\bar{j}}]^2 E_{\bar{i}\bar{l}}, & 
\bar{l}< \bar{i}< \bar{j} {\ \rm or \ } \bar{j}< \bar{i}< \bar{k},
\end{array}
\right.
$$
$$
[E_{\bar{i}\bar{j}},E_{\bar{k}\bar{j}}]=\left\{
\begin{array}{ll}
-E_{\bar{k}\bar{j}}, & \bar{k} < \bar{i} < \bar{j},\; 
\bar{j} < \bar{i} < \bar{k},\; \bar{k}\neq \bar{j}, \\
-[\hat{\bar{i}},\hat{\bar{j}}]^2  E_{\bar{k}\bar{j}}, & 
\bar{i}< \bar{j}< \bar{k} {\ \rm or \ } \bar{k}< \bar{j}< \bar{i},\\
-[\hat{\bar{i}},\hat{\bar{k}}]^2 E_{\bar{k}\bar{j}}, & 
\bar{i}< \bar{k}< \bar{j} {\ \rm or \ } \bar{j}< \bar{k}< \bar{i},
\end{array}
\right.
$$
$$
[E_{\bar{i}\bar{j}},E_{\bar{j}\bar{i}}]=[\hat{\bar{i}},\hat{\bar{j}}]^2
(E_{\bar{i}\bar{i}}-E_{\bar{j}\bar{j}}),
$$
$$
[E_{ij},E_{j\bar l}]=\left\{
\begin{array}{ll}
(i,j)^2E_{i\bar l}, & i<j, \\
E_{i\bar l}, & i>j, \end{array} \right. \quad
[E_{ij},E_{\bar ki}]=\left\{
\begin{array}{ll}
-E_{\bar kj}, & i<j, \\
-(j,i)^2E_{\bar kj}, & i>j, \end{array} \right. ,
$$
$$
[E_{\bar i\bar j},E_{k\bar i}]=\left\{
\begin{array}{ll}
-E_{k\bar j}, & \bar i<\bar j, \\
-[\hat{\bar j},\hat{\bar j}]^2
E_{k\bar j}, & \bar i>\bar j, \end{array} \right. \quad
[E_{\bar i\bar j},E_{\bar jl}]=\left\{
\begin{array}{ll}
[\hat{\bar i},\hat{\bar j}]^2
E_{\bar il}, & \bar i<\bar j, \\
E_{\bar il}, & \bar i>\bar j, \end{array} \right. ,
$$
$$
\left\{ E_{i\bar j},E_{\bar jl}\right\}=
\left\{ \begin{array}{ll}
\nu^2[1,\hat{\bar j}]^2(1,i)^2E_{il}, & i<l, \\
\nu^2[1,\hat{\bar j}]^2(1,l)^2E_{il}, & i>l, \end{array} \right.
$$
$$
\left\{ E_{i\bar j},E_{\bar ki}\right\}=
\left\{ \begin{array}{ll}
\nu^2(1,i)^2[1,\hat{\bar j}]^2E_{\bar k\bar j}, & \bar j<\bar k, \\
\nu^2(1,i)^2[1,\hat{\bar k}]^2E_{\bar k\bar j}, & \bar j>\bar k, \end{array}
 \right.
$$
\begin{equation}
\left\{ E_{i\bar j},E_{\bar ji}\right\}=
\nu^2(1,i)^2[1,\hat{\bar j}]^2(E_{ii}+E_{\bar j\bar j}).
\label{scr}
\end{equation}

For $\nu=\iota$  superalgebra $sl(m|n)$
is contracted to inhomogeneous superalgebra, which is semidirect sum
$ \{E_{i\bar{j}}, E_{\bar{i}j} \} \S (sl(m) \bigoplus sl(n)),$
with all anticommutators of the odd  generators equal to zero.

\subsection{Example: CK contractions of $sl(2|1)$}

The generators of superalgebra $sl(2;j_1;\nu|1)$ is given by \cite{Fra}
$$
H=\left ( 
\begin{array}{cc|c}
{1 \over 2} & 0 & 0  \cr 
0 & -{1 \over 2} & 0  \cr \hline
0 & 0 & 0 \cr 
\end{array}
\right ), \quad
Z=\left ( 
\begin{array}{cc|c}
{1 \over 2} & 0 & 0  \cr 
0 & {1 \over 2} & 0  \cr \hline
0 & 0 & 1  \cr 
\end{array}
\right ), \;
$$
$$
E_{12}=E^+=\left ( 
\begin{array}{cc|c}
0 & j_1 & 0  \cr 
0 & 0 & 0  \cr   \hline
0 & 0 & 0 \cr 
\end{array}
\right ), \quad
E_{21}=E^-=\left ( 
\begin{array}{cc|c}
0 & 0 & 0  \cr 
j_1 & 0 & 0  \cr \hline
0 & 0 & 0 \cr 
\end{array}
\right ),
$$
$$
E_{13}=\bar{F}^+=\left ( 
\begin{array}{cc|c}
0 & 0 &  \nu  \cr 
0 & 0 & 0  \cr  \hline
0 & 0 & 0 \cr 
\end{array}
\right ),    \;
E_{31}=F^-=\left ( 
\begin{array}{cc|c}
0 & 0 & 0  \cr 
0 & 0 & 0  \cr  \hline
 \nu & 0 & 0 \cr 
\end{array}
\right ), \;
$$
\begin{equation}
E_{32}=F^+=\left ( 
\begin{array}{cc|c}
0 & 0 & 0  \cr 
0 & 0 & 0  \cr   \hline
0 & \nu j_1 & 0 \cr 
\end{array}
\right ), \;
E_{23}=\bar{F}^-=\left ( 
\begin{array}{cc|c}
0 & 0 & 0  \cr 
0 & 0 & \nu j_1  \cr \hline
0 & 0 & 0 \cr 
\end{array}
\right )
\label{gen}
\end{equation}
and acts on the superspace $(x_1,j_1x_2|\nu \theta_1).$
The commutation relations are represented as
$$
[H,E^{\pm}]=\pm E^{\pm} ,\;
[E^+,E^-]=2j_1^2 H,  \; [Z,H]=[Z,E^{\pm}]= 
[E^{\pm},\bar{F}^{\pm}]= [E^{\pm},F^{\pm}]=0 ,\;
$$
$$
[H,\bar{F}^{\pm}]=\pm {1 \over 2} \bar{F}^{\pm} ,
\; [H,{F}^{\pm}]=\pm{1 \over 2} {F}^{\pm},\;
[Z,F^{\pm}]={1 \over 2} F^{\pm}, \; [Z,\bar{F}^{\pm}]=-{1 \over 2} 
\bar{F}^{\pm},
$$
$$
[E^{+},F^{-}]=-F^{+}, \; 
[E^{-},F^{+}]=-j_1^2F^{-}, \; 
[E^{+},\bar{F}^{-}]=j_1^2\bar{F}^{+} , \; 
[E^{-},\bar{F}^{+}]=\bar{F}^{-} ,  
$$
$$
\{F^{+},\bar{F}^{-}\}= \nu^2 j_1^2 (Z - H), \quad
\{F^{-},\bar{F}^{+}\}=\nu^2 (Z + H),
$$
\begin{equation}
\{\bar{F}^{+},F^{+}\}=\nu^2E^{+}, \;
\{\bar{F}^{-},F^{-}\}=\nu^2E^{-}, \;
\{\bar{F}^{+},\bar{F}^{-}\}=
\{{F}^{+},{F}^{-}\}=0.
\label{screl}
\end{equation}

For $\nu=\iota$ we obtain the semidirect sum of the abelian
odd subalgebra with the direct sum of the even subalgebras, namely,
$sl(2;j_1;\iota|1)=\{F^{\pm},\bar{F}^{\pm}\} \S (u(1)\oplus sl(2)).$
Two-dimensional contraction $\nu=\iota, j_1=\iota_1$ give in result
similar semidirect sum
$ sl(2;\iota_1;\iota|1)=\{F^{\pm},\bar{F}^{\pm}\} \S 
(u(1)\oplus sl(2;\iota_1))$
but with the subalgebra $ sl(2;\iota_1)=\{H,E^{\pm}\}$ instead of $sl(2).$
Under contraction $j_1=\iota_1$ we have the semidirect sum
$sl(2;\iota_1;\nu|1)= \{E^{\pm}, F^+, \bar{F}^- \} \S
\{H, Z, F^-, \bar{F}^+ \} $ of the subsuperalgebras each of them 
generated both even and odd generators.

\section{Casimir operators}

The study of Casimir operators plays a grate role in the 
representation theory of simple Lie algebras since their
eigenvalues characterize a representations. In the case
of Lie superalgebras their eigenvalues completely characterize
a typical representation while they are identically vanishing
on an atypical representation. 
An element $C$ of universal enveloping superalgebra $U(A)$ commuting
with all elements of $U(A)$ is called a Casimir operator of
superalgebra $A.$ The algebra of the Casimir operators of $A$
is the $Z_2$-center of $U(A).$

 Casimir operators of the basic Lie superalgebras
can be constructed as follows \cite{Fra}, \cite{LS}, \cite{ACF}. 
Let $A=sl(m|n)$ with $m\neq n$ or
$osp(m|n)$ be a basic Lie superalgebra. Let $\{E_{IJ}\}$ be 
a matrix basis of generators of $A$ where $I,J=1,\ldots,m+n$
with ${\rm deg}I=0$ for $I=1,\ldots,m$ and ${\rm deg}I=1$ for 
$I=m+1,\ldots,m+n.$
Then defining $(\bar{E})_{IK}=(-1)^{{\rm deg}K}E_{IK},$ a standard sequence
of Casimir operators is given by 
$$
C_p=str(\bar{E}^p)=\sum_{I=1}^{m+n} (-1)^{{\rm deg}I}(\bar{E}^p)_{II}=
$$
\begin{equation}
=\sum_{I,I_1,\ldots,I_{p-1}=1}^{m+n}E_{II_1}(-1)^{{\rm deg}I_1}\ldots
E_{I_kI_{k+1}}(-1)^{{\rm deg}I_{k+1}}\ldots E_{I_{p-1}I}.
\label{6.1}
\end{equation}
In the case of $sl(m|n)$ with $m \neq n$ one finds for example
$ C_1 =0$ and
\begin{equation}
C_2= \sum_{i,j=1}^{m}E_{ij}E_{ji}-
\sum_{\bar{k},\bar{l}=m+1}^{m+n}E_{\bar{k}\bar{l}}E_{\bar{l}\bar{k}}
+\sum_{i=1}^{m}\sum_{\bar{k}=m+1}^{m+n}(E_{\bar{k}i}E_{i\bar{k}}-
E_{i\bar{k}}E_{\bar{k}i})-{m-n \over mn} Y^2.
\label{6.2}
\end{equation}
The diagonal elements of matrix $\bar{E}$ are taken in the form
$ (\bar{E})_{ii}=E_{ii}+{{1}\over {m}}Y, \;
(\bar{E})_{\bar{k}\bar{k}}=-E_{\bar{k}\bar{k}}+{{1}\over {n}}Y$
and two conditions on generators:
$\sum_{i=1}^{m}E_{ii}=0,\;
\sum_{\bar{k}=m+1}^{m+n}E_{\bar{k}\bar{k}}=0$ 
are taken into consideration.
In the case of $osp(m|n)$  one finds 
$ C_1 =0$ and
\begin{equation}
C_2= \sum_{i,j=1}^{m}E_{ij}E_{ji}-
\sum_{\bar{k},\bar{l}=m+1}^{m+n}E_{\bar{k}\bar{l}}E_{\bar{l}\bar{k}}
+\sum_{i=1}^{m}\sum_{\bar{k}=m+1}^{m+n}(E_{\bar{k}i}E_{i\bar{k}}-
E_{i\bar{k}}E_{\bar{k}i}).
\label{6.3}
\end{equation}

One has to stress that unlike the algebraic case, the center of
$ U(A)$ for the classical Lie superalgebras is in general not finitely
generated. For only Lie superalgebra $osp(1|2n)$ the center of its
universal enveloping superalgebra
is generated by $n$ Casimir operators of degree $2,4, \ldots ,2n.$

To obtain Casimir operators of superalgebra $sl(m;j|n;\epsilon)$
we shall proceed in the standart manner. First we get the matrix
$\bar{E}(j;\epsilon).$ For this we put in matrix $\bar{E}$ the new
generators of $sl(m;j|n;\epsilon)$ instead of the old ones of $sl(m;n)$
according to (\ref{5.1+}) and denote the obtained matrix 
as $\bar{E}(\rightarrow).$ In general its elements are undefined
for nilpotent values of parameters $j,\epsilon,\nu.$ 
So it is necessary to multiply $\bar{E}(\rightarrow)$ on minimal
multiplier which eliminate all undefined expressions  in matrix 
elements, namely,
$\nu(1,m)[1,n].$  Finally we have
\begin{equation}
\bar{E}(j;\epsilon)=
\nu(1,m)[1,n]\bar{E}(\rightarrow)
\label{6.4}
\end{equation}
with matrix elements $(k\neq p, \; \bar{k} \neq \bar{p})$
$$
(\bar{E}(j;\epsilon))_{kk} = \nu(1,m)[1,n](E_{kk}+{{1}\over{m}}Y), \quad
(\bar{E}(j;\epsilon))_{\bar{k} \bar{k}}= 
\nu(1,m)[1,n](-E_{\bar{k}\bar{k}}+{{1}\over{n}}Y), 
$$
$$
(\bar{E}(j;\epsilon))_{kp} = \nu(1,k)(p,m)[1,n]E_{kp}, \quad
(\bar{E}(j;\epsilon))_{\bar{k} \bar{p}}= \nu(1,m)[1,\hat{\bar{k}}]
[\hat{\bar{p}},n]E_{\bar{k}\bar{p}}, 
$$
\begin{equation}
(\bar{E}(j;\epsilon))_{i \bar{k}}= -(i,m)[\hat{\bar{k}},n]E_{i\bar{k}},\quad
(\bar{E}(j;\epsilon))_{\bar{i}k}= (k,m)[\hat{\bar{i}},n]E_{\bar{i}k}.
\label{6.5}
\end{equation}
Maximal multiplier $\nu(1,m)[1,n]$ have the diagonal elements
and minimal unit multiplier have the matrix elements
$
(\bar{E}(j;\epsilon))_{m,m+n} = E_{m,m+n}, \;
(\bar{E}(j;\epsilon))_{m+n,m} = E_{m+n,m}.
$

The sequence of Casimir operators of $sl(m;j|n;\epsilon)$
is given by
\begin{equation}
C_p(j;\epsilon)={\rm str}\bar{E}^p(j;\epsilon)=
\nu^p(1,m)^p[1,n]^p {\rm str} (\bar{E}(\rightarrow))^p.
\label{6.6}
\end{equation}
Indeed, let $X^{\star}$ be an arbitrary generator of $sl(m|n)$.
Under computing $[C_p,X^{\star}]=0$ we get identical terms 
but with different signs (plus and minus) so their sum is equal to zero.
Under transformation of this commutator to the corresponding commutator
of $sl(m;j|n;\epsilon)$ identical terms are multiplied on identical
multipliers therefore their sum remains equal to zero, i.e.
$[C_p(j;\epsilon),X]=0.$


Let us illustrate the above expressions on the simple
example of $sl(2;j_1|1)$ superalgebra. The generators are
transformed as follows
$$
E_{11}=E_{11}^{\star},\; Y= Y^{\star}, \;
E_{12}=j_1E_{12}^{\star},\; 
E_{21}=j_1E_{21}^{\star},\; 
E_{13}=\nu E_{13}^{\star},\; 
$$
\begin{equation}
E_{31}=\nu E_{31}^{\star},\; 
E_{23}=\nu j_1E_{23}^{\star}, \;
E_{32}=\nu j_1E_{32}^{\star}.
\label{6.7}
\end{equation}
and matrix $\bar{E}(j_1)$ according to (\ref{6.4}),(\ref{6.5})
is given by
$$
\bar{E}(j_1)=\nu j_1\bar{E}(\rightarrow)=
\nu j_1 \left(
\begin{array}{cc|c}
E_{11}+{1 \over 2} Y & {1 \over j_1}E_{12} &-{1\over \nu}E_{13} \cr
{1 \over j_1}E_{21} & -E_{22}+{1 \over 2}Y &-{1\over \nu j_1} E_{23} \cr \hline
{1 \over \nu}E_{31} & {1\over \nu j_1} E_{32} & Y \cr
\end{array}
\right)=
$$
\begin{equation}
= \left(
\begin{array}{cc|c}
\nu j_1(E_{11}+{1 \over 2}Y) & \nu E_{12} &-j_1E_{23} \cr
\nu E_{21} & \nu j_1(-E_{22}+{1 \over 2}Y) &-E_{23} \cr \hline
j_1E_{31} &  E_{32} & \nu j_1 Y \cr
\end{array}
\right).
\label{6.8}
\end{equation}
The first order Casimir operator disappear
$C_1(j_1)={\rm str}\bar{E}(j_1)=0.$
The second order Casimir operator is as follows
$$
C_2(j_1)={\rm str} (\bar{E}(j_1))^2=
\nu^2 j_1^2\left(2E^2_{11}-{1\over 2} Y^2 \right)+
\nu^2 \left(E_{12}E_{21}+E_{21}E_{12} \right)+
$$
\begin{equation}
+j_1^2 \left(E_{31}E_{13}+E_{13}E_{31} \right)+
E_{32}E_{23}-E_{23}E_{32}.
\end{equation} 

In the case of superalgebras $osp(M|N)$ the  multiplier in (\ref{6.5})
is equal to $\nu (1,M)[1,{N \over 2}]$
and all formulas for matrix $\bar{E}(j;\epsilon)$
and matrix elements $\left(\bar{E}(j;\epsilon)\right)_{kp}$ 
appear as for the $sl(m;j|n;\epsilon)$ with substitution $m=M$ and 
$n={N \over 2}$.
Let us consider the $osp(1|2;\nu)$ superalgebra as an example. 
Their generators are transformed as
\begin{equation}
E_{12}=\nu E_{12}^{\star},\; 
E_{13}=\nu E_{13}^{\star},\; 
E_{23}= E_{23}^{\star}, \;
E_{32}= E_{32}^{\star}, \;
E_{22}= E_{22}^{\star},
\end{equation}
and matrix $\bar{E}(\nu)$ is given by
\begin{equation}
\bar{E}(\nu)=
-\nu  \left(
\begin{array}{c|cc}
0                & {1 \over \nu }E_{12} & {1\over \nu }E_{13} \cr \hline
{1 \over \nu }E_{13} & E_{22} & E_{23} \cr
-{1 \over \nu }E_{12} &  E_{32} & -E_{22} \cr
\end{array}
\right)
=- \left(
\begin{array}{c|cc}
0   & E_{12}     &      E_{13} \cr \hline
E_{13}  & \nu E_{22} &  \nu E_{23} \cr 
- E_{12}& \nu E_{32} & -\nu E_{22} \cr
\end{array}
\right).
\label{6.11}
\end{equation}
The first order Casimir operator is equal to zero
$C_1(\nu)={\rm str}\bar{E}(\nu)=0$
and the second order Casimir operator is represented as
\begin{equation}
C_2(\nu)=
\nu^2 E^2_{22}+
\left(E_{12}E_{13}-E_{13}E_{12} \right)-
{1 \over 2}\nu^2 \left(E_{32}E_{23}+E_{23}E_{32}\right).
\end{equation}

\section{Conclusion}

 Using classical CK Lie algebras of different type 
we have built basic CK superalgebras. Unlike standard procedure
\cite{IW} of zero tending parameter contractions in this work 
are described with the help of nilpotent valued parameters. 
Such approach gives an opportunity to obtain the distribution of 
contraction parameters among superalgebra generators starting
from quadratic form and hence to build CK superalgebras by means 
of pure algebraic tools without limiting procedure. 
Contracted superalgebras are connected with transformations of 
superspaces with nilpotent cartesian coordinates and represent 
a wide class of different semidirect sums for different possible 
contractions. An infinite sequences of Casimir elements of CK 
superalgebras have been obtained by a suitable transformations of 
the standard expressions of the corresponding operators of the basic 
superalgebras. It is our hope that CK superalgebras will be relevant 
for construction of supersymmetric physical models.

\section*{Acknowledgments}
This work was supported by Russian Foundation for Basic
Research under Project 01-01-96433.



\begin{thebibliography}{99}

\bibitem{1}
 Golfand Yu A and  Likhtman E P 1971 {\it 
 JETP. Lett.}{\bf 13} 452

\bibitem{2}
 Volkov D V and  Akulov V P 1972 {\it JETP Lett.}
{\bf 16} 621

\bibitem{3}
  Wess J and  Zumino B 1974 {\it Nucl. Phys.}
{\bf B70} 139 

\bibitem{W-86}
 West P  1986 {\it Introduction to supersymmetry and supergravity},
(Singapore: World Scientific)

\bibitem{K-75}
 Keck B W 1975 {\it J.Phys.A: Math.Gen.}
{\bf 8} 1819

\bibitem{Vas-90}
 Vasiliev M A 1990 {\it Phys.Lett.B}
{\bf 243} 378

\bibitem{B-96}
 Bars I 1996 {\it Preprint} hep-th/9608061

\bibitem{BIK}
 Bellucci S,  Ivanov E and  Krivonos S 2001
{\it Nucl.Phys.B (Proc. Suppl.)} {\bf 102\&103} 26 

\bibitem{NG}
Gromov N A 1990 {\it Contractions and Analytical Continuations of
Classical Groups. Unified Approach} (Syktyvkar: Komi SC) (in Russian)

\bibitem{JMP}
Gromov N A and Man'ko V I 1990 {\it J.Math.Phys.} {\bf 31} 1054, 1060



\bibitem{Rem}
 Rembielinski J, Tybor W   1984
{\it Acta Physica Polonica} {\bf B15} 611

\bibitem{Val-99}
 Hussin V,  Negro J and  del Olmo M A 1999 {\it J.Phys.A: Math.Gen.}
{\bf 32} 5097 

\bibitem{Pat}
 Patra M K and  Tripathy K C  1989
{\it Lett.Math.Phys} {\bf 17} 1 

\bibitem{MP}
Moody R V and Patera J 1991 {\it J.Phys.A: Math.Gen.}{\bf 24} 2227

\bibitem{GKK}
Gromov N A, Kostyakov I V and Kuratov V V
{\it Preprint} arXiv:hep-th/0110257

\bibitem{Fra}
 Frappat L,  Sciarrino A and  Sorba P 1996
{\it Dictionary on Lie Superalgebras}
(hep-th/9607161, ENSLAPP-AL-600/96 and DSF-T-30/96)

\bibitem{LS}
 Leites D and  Sergeev A 2002
{\it Preprint} arXiv:math.RT/0202180 v1

\bibitem{ACF}
 Arnaudon D,  Chryssomalakos C and  Frappat L 1995
{\it Preprint} arXiv:q-alg/9503021 v2, ENSLAPP-A-505/95

\bibitem{IW}
In{\"o}n{\"u} E and Wigner E P 1953
{\it Proc.Nat.Acad.Sci. USA} {\bf39} 510


\end{thebibliography}
\end{document}